\documentstyle[aps]{revtex}   
\begin{document}
\draft
\title{Weyl-type Fields with Geodesic Lines of Force}
\author{Brendan S. Guilfoyle \footnote{electronic mail: brendan.guilfoyle@ittralee.ie}}
\address{Mathematics Department, Institute of Technology Tralee, Tralee, Co. Kerry, Ireland.}
\date{\today}
\maketitle
\begin{abstract}
The static electrogravitational equations are studied and it is shown that an aligned type D metric which has a Weyl-type relationship between the gravitational and electric potential has shearfree geodesic lines of force.  All such fields are then found and turn out to be the fields of a charged sphere, charged infinite rod and charged infinite plate.  A further solution is also found with shearing geodesic lines of force.  This new solution can have $m>|e|$ or $m<|e|$, but cannot be in the Majumdar-Papapetrou class (in which $m = |e|$). It is algebraically general and has flat equipotential surfaces.
\end{abstract}

\newtheorem{thm}{Theorem}[section]   
\newtheorem{cor}[thm]{Corollary}   
\newtheorem{lem}[thm]{Lemma}   
\newtheorem{prop}[thm]{Proposition}   
\newtheorem{defn}{Definition}[section]      
\newtheorem{rem}{Remark}[section]   
\newcommand{\nc}{\newcommand}
\nc{\sigf}{\Sigmagma_\phi}
\nc{\bg}{\begin{eqnarray}}
\nc{\ed}{\end{eqnarray}}
\nc{\bp}{\bar{\Psi}}
\nc{\bm}{\bar{m}}
\nc{\ul}{\underline}
\nc{\bee}{\begin{equation}}
\nc{\ee}{\end{equation}}
\nc{\beer}{\begin{eqnarray}}
\nc{\eer}{\end{eqnarray}}
\nc{\M}{$(M,g_{ab})\:$}
\nc{\V}{$(V^3,g_{ij})\:$}
\nc{\dv}{\textstyle\frac}
\nc{\sdv}{\scriptstyle\frac}
\nc{{{\hf}}}{\dv{1}{2}}
\nc{\qtr}{\dv{1}{4}}
\nc{{{\shf}}}{\sdv{1}{2}}
\nc{\nn}{\nonumber}
\nc{\2}{\;\;}
\nc{\3}{\;\;\;}
\nc{\4}{\;\;\;\;}
\nc{\5}{\;\;\;\;\;}
\nc{\6}{\;\;\;\;\;\;}
\nc{\tr}{$\la_{(A)}^{\quad i}\:$}
\nc{\tn}{$\la_{(1)}^{\quad i}\:$}
\nc{\tw}{$\la_{(2)}^{\quad i}\:$}
\nc{\tth}{$\la_{(3)}^{\quad i}\:$}
\nc{\tb}{\bar{2}}
\nc{\ds}{\parallel}
\nc{\VB}{$(\overline{V}^3,\overline{g}_{ij})\:$}
\nc{\lap}{\bigtriangleup_2}
\nc{\lapb}{\overline{\bigtriangleup}_2}
\nc{\len}{\bigtriangleup_1}
\nc{\lenb}{\overline{\bigtriangleup}_1}
\nc{\imp}{$\Rightarrow$}
\nc{\dimp}{$\Leftrightarrow$}
\nc{\om}{\omega}
\nc{\de}{\delta}
\nc{\al}{\alpha}
\nc{\be}{\beta}
\nc{\ep}{\epsilon}
\nc{\ga}{\Gamma}
\nc{\la}{\lambda}
\nc{\La}{\Lambda}
\nc{\si}{\sigma}
\nc{\Si}{\Sigma}
\nc{\ka}{\kappa}
\nc{\et}{\eta}
\nc{\ptl}{\partial}
\nc{\oa}{$\om,_{\al}$}
\nc{\cdb}{\overline{\bigtriangledown}}
\nc{\gt}{\rightarrow}

\section*{I. Introduction}
Static fields have been a very fruitful area of study for general relativity. On the one hand, the simplifications in the field equations for static fields are substantial, and on the other hand, it is this area that lends itself most easily to comparison with Newtonian gravitation and classical potential theory. For example, the 44-component of the metric tensor, $g_{44}$, plays the role of a gravitational potential. This allows one to talk of gravitational  equipotential surfaces (surfaces of constant $g_{44}$) and lines of force (the integral curves of the gradient of $g_{44}$).

Das \cite{das} and Kota and Per\c{j}es \cite{kap} investigated vacuum metrics in which the lines of force are geodesic. The Newtonian analogue for such a field is one with straight lines of force, which is generated by a sphere or an infinite rod or an infinite plate. Das found that amongst all vacuum fields with geodesic lines of force is included the exterior field of a sphere and an infinite rod and an infinite plate. These constitute all of the vacuum fields in which the lines of force are in addition shearfree.  Das also discovered a shearing field with no Newtonian analogue. 

 When the source of the field is an electric field, one also has the electric potential $\phi$, and the corresponding electric equipotential surfaces and lines of force. Fields in which there is a functional relationship between the two potentials has been investigated by numerous authors (e.g. Weyl \cite{weyl}, Majumdar \cite{maj}, Papapetrou \cite{pap1}, Gatreau and Hoffman \cite{gah}, Guilfoyle \cite{guilgrg}). We will refer to such fields as being of {\it Weyl-type}.

In this paper we show that amongst all Weyl-type fields the algebraically special (type D) ones have shearfree geodesic lines of force. We then go on to find all Weyl-type fields with geodesic lines of force. The shearfree metrics turn out to describe the fields of a charged sphere, an infinite charged rod and an infinite charged plate. The only asymptotically flat member of this class is the Reissner-Nordstr\"{o}m  metric.  Again, there is a further shearing field which is algebraically general and has no Newtonian analogue. We find, however, that this is not in the Majumdar-Papapetrou class of electrovac solutions. In fact, the only Majumdar-Papapetrou field with geodesic lines of force is found to be the extreme Reissner-Nordstr\"{o}m solution.

\section*{II. Background and General Results}
A space-time $(M,g_{\mu\nu})$ is {\it static} if there exists a timelike hypersurface orthogonal Killing vector $\xi^\mu$ on $M$.  With a suitable choice of co-ordinates $(x^i, t)$ the metric takes the form
\[ds^2 = g_{ij}dx^idx^j-e^\omega dt^2,\]
where both the spatial metric $g_{ij}$ and $e^\omega$ depend only on $x^i$. Here, and throughout, Greek letters will take values 1 to 4, while Latin letters will take values 1,2,3. We will raise and lower all indices using $g$.

The 3+1 split above induces the following decomposition of the Riemann tensor 
of ($M,g_{\mu\nu}$) in terms of that of the spatial slices ($V^3,g_{ij}$)  (see Synge \cite{synge}):
\begin{eqnarray}
R_{ijkl} & = & ^3\!R_{ijkl} \nn \\
R_{i44l} & = & -e^{{{\shf}}\omega}(e^{{{\shf}}\omega})_{|il}\label{e:riemred} \\
R_{ij4l} & = & 0. \nn
\end{eqnarray} 
A stroke represents covariant differentiation with respect to the Levi-\c{C}ivita connection of the spatial metric $g_{ij}$.
The Ricci tensors of ($M,g_{\mu\nu}$) and ($V^3,g_{ij}$) are then found to be related by 
\begin{eqnarray}
R_{ij} & = & ^3\!R_{ij}  +  e^{-{\shf}\omega}(e^{{\shf}\omega}) _{|ij} \label{e:riccired1} \\
R_{44} & = & -{\hf} e^{\omega}(\triangle\omega + {\hf}|d\omega|^2), \label{e:riccired2}
\end{eqnarray}
where $\triangle$ is  the covariant laplacian $\triangle\omega = g^{ij}\omega_{|ij}$ and $|d\omega|^2 = g^{ij}\omega,_i\omega,_j$.

We note that the 3-dimensional Riemann tensor of ($V^3,g_{ij}$) is entirely determined 
by the 3-dimensional Ricci tensor through
\begin{equation}\label{e:3driem}
^3\!R_{ijkl}  =  g _{i[l}\:^3\!R _{k]j}  +  
g _{j[k}\:^3\!R _{l]i} +  {\hf}\:^3\!Rg_{i[k}\:g_{l]
j},
\end{equation}
where $^3\!R$ is the Ricci scalar \(^3\!R \equiv \:^3\!R^{i }_{i}\). Skew-symmetrisation (symmetrisation) is denoted by square (round) brackets on pairs of indices.

It is often useful to turn to the conformally related background space \VB given by
\[g_{ij} = e^{-\omega}\;\overline{g}_{ij}.\]
The relationship between the Ricci curvatures of \V and \VB is
\begin{equation}
^3\!R_{ij}  =  \!^3\:\overline{R}_{ij} - e^{-\shf \omega}(e^ {\shf
 \omega})_{|ij}  + {\hf} \omega,_{i}\omega,_{j} -{\hf} 
g_{ij} (\triangle \omega + {\hf} |d\omega|^2).
\end{equation}
Einstein's field equations for the electric field $\phi$ are
\beer 
R_{ij}& = &2e^{-\omega}\phi,_{i}\phi,_{j}  -e^{-\omega}|d\phi|^2g_{ij}\label{e:ein1}\\
\triangle\omega& = & 2e^{-\omega} |d\phi|^2 -{{\hf}}|d\omega|^2 \label{e:ein2}\\
R& = &0 ,\label{e:ein3}
\eer
while Maxwell's equations reduce to
\bee
\triangle\phi = {{\hf}}\omega^{,i}\phi,_{i}.\label{e:max}
\ee

\section*{III. Algebraic Structure}

We now turn to the algebraic structure of the electrostatic field. 
The most widely used classification scheme for the gravitational field 
is due to Petrov \cite{pet} (see also Kramer {\it et al.} \cite{kram}, Papapetrou \cite{pap2} and Synge \cite{synge1}).  In this, a metric is classified by 
the degeneracy of certain null directions associated with the Weyl conformal curvature tensor. This tensor is given in 
terms of the Riemann and Ricci curvature tensors by
\[C_{\alpha\beta\gamma\delta}  =  R_{\alpha\beta\gamma\delta}  + {\hf}(g_{\alpha[\gamma}R_{\delta]\beta} + g_{\beta[\delta}R_{\gamma]\alpha})  +  {\dv{1}{6}}R
g_{\alpha[\delta}g_{\gamma]\beta}. \]
A {\em principal null direction} ({\em p.n.d.} for short) of the Weyl 
tensor is a null vector, $k^\mu$, satisfying
\[
k_{[\mu}C_{\alpha]\beta\gamma[\delta}k_{\nu]}k^\beta k^\gamma  =  0. 
\]
As the name suggests, a p.n.d. is defined only up to multiplication 
by a function.

A {\it repeated} p.n.d. for the Weyl tensor is a null vector, $k^\mu$, satisfying
\[
C_{\alpha\beta\gamma[\delta}k_{\nu]}k^\beta k^\gamma=0.
\]

Using  equations (\ref{e:riemred}) to  (\ref{e:ein3}) this can be shown to be equivalent to 
\bee\label{e:con3}
e^{-{\shf}\omega}(e^{{\shf}\omega})_{|ij}  =  e^{-\omega}\phi,_{i}\phi,_{j} - 
  \chi\left(\frac{k_{i}k_{j}}{ |k|^2} - \dv{1}{3}g_{ij}\right),
\ee
where 
\bee\label{e:con3a}
\chi \equiv \frac{3}{2 |k|^2}[e^{-\omega}(\;k^i\phi,_i\; )^2 - e^{-{\shf}\omega}(e^{{\shf}\omega})
  _{|ij}k^{i}k^{j}],
\ee
and the p.n.d. is
\bee\label{e:pnd}
 k^\mu  =  (k^{i},e^{-{\shf}\omega}|k|). 
\ee
We note as a check on these calculations that contracting (\ref{e:con3}) gives 
\[ e^{-{\shf}\omega}\triangle(e^{{\shf}\omega})  =  e^{-\omega}|d\phi|^2 \]
which is precisely the field equation (\ref{e:ein2}).

Thus a necessary and sufficient condition that a 
static electrovac space-time is algebraically special (and therefore, type D) is that there exists a 3-vector $k^{i}$ 
satisfying (\ref{e:con3}) and (\ref{e:con3a}). Then the repeated p.n.d. of the Weyl tensor is given by (\ref{e:pnd}) and the Ricci tensor of \V is (put (\ref{e:con3}) in (\ref{e:riccired1}))
\bee\label{e:con4}
^3\!R_{ij}  =  \chi\left(g_{ij}-\frac{3k_{i}k_{j}}{|k|^2}\right) +  
  e^{-\omega}(\phi,_{i}\phi,_{j} - |d\phi|^2\: g_{ij}).
\ee

In the vacuum case ($\phi = 0$) this tells us that a repeated p.n.d.
 of the Weyl tensor is also an eigenvector of the spatial Ricci tensor 
$^3\!R_{ij}$. Furthermore, any vector on \V  which is orthogonal to 
the spatial component, $k^{i}$, of the repeated p.n.d., is also an 
eigenvector of the spatial Ricci tensor. This decomposition of $^3\!R_{ij}$ 
allows one to integrate the field equations fully, and determine explicitly all 
type D static vacuum fields, as was done almost 75 years ago by 
Levi-\c{C}ivita \cite{lc}.  Although the Schwarzchild solution is of 
this type, it was found that all the other 6 solutions in this class are 
unphysical (Kramer {\it et al.} \cite{kram}, Ehlers and Kundt \cite{eak}).

If we now include the electromagnetic field, the spatial Ricci tensor differs 
only by the projection operator onto directions orthogonal to $\phi,_{i}$.
Thus we have the following theorem:

\begin{thm}
 A repeated p.n.d. of the Weyl conformal tensor of an algebraically special 
static electrovac field is also an eigenvector of the spatial Ricci tensor iff 
one of the following holds:
\begin{description}
\item{(i)}
The electric field vanishes ($\phi = 0$).
\item{(ii)}
The electric field is aligned ($\phi,_{[i}k_{j]} = 0$).
\item{(iii)}
The electric field is perpendicular ($\phi,_{i}k^{i} = 0$).
\end{description}
\end{thm}

Since the field equations are tractable in case {\it (i)} and the algebraic 
structure of $^3\!R_{ij}$ is the same in {\it (i)}, {\it (ii)} and {\it (iii)}, we are led 
to consider the possibility that all aligned or perpendicular type D static 
electrovac fields may be explicitly determined using equation (\ref{e:con4}) to 
integrate the field equations.

We shall look at the case where the field is aligned and $\omega,_{i}$ 
is also coincident with $\phi,_{i}$. In this case, there is a functional 
relationship between $e^{\omega}$ and $\phi$. It is well known that this 
must be of the form (see Majumdar \cite{maj})
\[e^{\omega}  =  A  +  B\phi  +  \phi^2,\]
from which we find that 
\[\phi,_{i}  =  \frac{e^{\omega}}{2(e^{\omega} + \lambda)^{{\shf}}} \omega,_{i},\]
and
\[
|d\phi|^2  = \frac{e^{2\omega}|d\omega|^2}{4(e^{\omega} + \lambda)},
\]
where 
\[\lambda \equiv \dv{B^2}{4} - A \qquad A,B\quad constants. \]
From these we find that (\ref{e:con3}) becomes
\bee\label{e:con5}
\omega_{|ij}  =  -\frac{\lambda}{2(e^{\omega} + \lambda)}\omega,_{i}\omega,_{j}  +  
  2\chi\left({\dv{1}{3}}g_{ij} - \frac{\omega,_{i}\omega,_{j}}{|d\omega|^2}\right),
\ee
or transvecting with $\omega^{,j}$
\[\omega_{|ij}\omega^{,j}  = -\left[{\dv{4}{3}}\chi  +  \frac{\lambda|d\omega|^2}{2(e^{\omega} + \lambda)}\right]
  \omega,_{i}.\]
This is just the geodesic equation,
\[ \omega_{|ij}\omega^{,j}  =  \frac{\omega_{lk}\omega^{,l}\omega^{,k}}
   {|d\omega|^2}\omega_{,i}\;.\]
We have therefore established the following theorem:

\begin{thm}
If a static Weyl-type electrovac space is aligned and type D, then the lines of force are geodesic in \V.
\end{thm}
        
In fact, we can go further by looking at these equations in the conformal space \VB. There, equation (\ref{e:con5}) becomes: 
\[\omega_{\|ij}  = \left[ -\frac{e^{\om}}{4(e^{\omega} + \lambda)}  + \frac{3}{2}\frac{\omega^{\|kl}\omega,_k\omega,_l} {\|d\omega\|^4}\right]\omega,_{i}\omega,_{j}  +  \left[\frac{\|d\om\|^2e^\om}{4(e^{\omega} + \lambda)} -\frac{1}{2}\frac{\omega^{\|kl}\omega,_k\omega,_l}{\|d\omega\|^2} \right] \overline{g}_{ij},
\]
where we have denoted covariant differentiation with respect to the metric $\overline{g}_{ij}$ by a double stroke subscript and $\|d\omega\|^2 = \overline{g}^{ij}\omega_i\omega_j$.
Thus, the lines of force are also geodesic in \VB. Now, if we let
\[k_i = \frac{\omega,_i}{\|d\omega\|},\]
then $k^i$ is a geodesic in \VB and $k^\mu$ (given by (\ref{e:pnd})) is also a null geodesic in the full space-time  ($M,g_{\mu\nu}$).  In addition, $k^i$ is shearfree in \VB:
\[k_{(i\|j)}k^{i\|j}-(\overline{g}^{ij}k_{i\|j})^2 = 0.\]
This also means, as one would expect from the Goldberg-Sachs theorem \cite{gs}, that that the p.n.d. $k^\mu$ is shearfree in ($M,g_{\mu\nu}$). Thus we have the following:

\begin{thm}
If a static Weyl-type electrovac space is aligned and type D, then the lines of force are geodesic and shearfree in \VB.
\end{thm}

In the next section we determine all such metrics explicitly.

\section*{IV. Type D Fields }

It is well known that the Reissner-Nordstr\"{o}m solution is the unique 
static asymptotically flat electrovac field with geodesic shearfree lines of 
force (or {\em eigenrays})  Kota and Per\c{j}es \cite{kap}. In this section we present all static electrovac fields with $g_{44} = g_{44}(\phi)$ which possess shearfree geodesic lines of force.

In the vacuum case, all static fields with geodesic lines of force were found 
by Das \cite{das} and Kota and Per\c{j}es \cite{kap}. Here we shall adopt the formalism of Das, generalizing it to include the electrostatic field. This formalism 
entails the setting up of a 3-dimensional complex triad field in \VB and then 
expressing the field equations in terms of the associated 3-dimensional Ricci 
rotation co-efficients.

Let  $\lambda_{(A)}^{\quad i}$ be an orthonormal frame in \VB and 
\[
\Gamma_{(ABC)} \equiv \lambda_{(A) j\| k}\lambda_{(B)}^{\quad j}\lambda_{(C)}^
{\quad k},
\]
be the associated Ricci rotation co-efficients. Here and throughout we use bracketed capital Latin letters $A, B$.. = 1,2,3 to represent frame components and summation is implied over any repeated indices. These satisfy the commutation relations
\bee\label{a1}
\la_{(A)}^{\quad i},_{ j}\la_{(B)}^{\quad j} - \la_{(B)}^{\quad i},_{ j}
  \la_{(A)}^{\quad j}  +  \ga_{(C[AB])}\la_{(C)}^{\quad i}  = 0 ,
\ee
and the Riemann curvature of \VB is given in terms of them by
\beer
^3\overline{R}_{(ABCD)}& = &\ga_{(AB)[(C),(D)]}\;  + \; \ga_{(ABM)}\ga_{(M[CD])}
  \;  + \;\ga_{(MAD)}\ga_{(MBC)} \nn\\
  & & \quad -\quad \ga_{(MAC)}\ga_{(MBD)}\label{aa1}.
\eer
We choose the frame so that the congruence of \tn is normal to 
the surface $\om = constant$. Thus
 \bee\label{a2}
\lambda_{(1) i}  =  U\om,_{ i} \qquad \Gamma_{(123)}  =  \Gamma_{(132)}.
\ee
We choose co-ordinates so that
 \[
\om = x^1 \qquad \overline{g}_{12} = \overline{g}_{13} = 0 .
\]
Hence (\ref{a2}) can be raised to give
 \[
\lambda_{(1)}^{\quad i} = U^{-1}\de_1^{ i}.
\]
Since the lines of force are geodesic we have that
 \[
\ga_{(131)} = \ga_{(121)} = 0.
\]
Parallely propagate \tw, \tth along \tn so that 
 \[
\ga_{(231)} = 0.
\]
Finally, since we have vanishing shear
 \[
\ga_{(122)} = \ga_{(133)} \qquad \ga_{(123)} = 0.
\]
Label the remaining non-vanishing rotation co-efficients by
\beer
H &\equiv& \dv{1}{2}(\Gamma_{(122)} + \Gamma_{(133)}) \nn\\
\sigma &\equiv& \dv{1}{\sqrt{2}}(\Gamma_{(233)} - i\Gamma_{(232)}). \nn
\eer
Here and throughout capital Latin letters will indicate real valued functions 
and Greek letters will denote complex valued functions. 

Furthermore, we make a formal transformation to complex conjugate 
co-ordinates defined by
 \[
z^2 \equiv x^2 + ix^3 \qquad z^{\tb} \equiv x^2-ix^3.
\]
It should be noted that a co-ordinate transformation of the form
 \beer
x^1 \gt x'^1 & = & x^1 \nn\\
z^2 \gt z'^2 & = & f(x^1,z^2) \nn\\
z^{\tb} \gt z'^{\tb} & = & \overline{f(x^1,z^2)} ,\nn
\eer
where f($x^1,z^2$) is an analytic function of $x^1,z^2$ and the bar denotes 
complex conjugation, does not alter the static form of the metric. This corresponds to a co-ordinate transformation of ($x^2,x^3$) on the 
equipotential surfaces $\om = x^1 = constant$.

The field equations written in terms of frame components in \VB are 
 \bee\label{feq1}
^3\!\overline{R}_{(AB)}  =  2e^{-\om}\phi,_{(A)}\phi,_{(B)} -{\hf}\om,_{(A)}\om,_{(B)} 
\ee
 \bee\label{feq2}
\om,_{(AA)}  +  \Gamma_{(CAA)}\om,_{(C)}  =  2e^{-\om}\phi,_{(A)}\phi,_{(A)}\; ,
\ee
with Maxwell's equation
 \bee\label{max}
\phi,_{(AA)}  +  \Gamma_{(CAA)}\phi,_{(C)}  =  \om,_{(A)}\phi,_{(A)}\; .
\ee
Finally, introduce the complex triad field \(\La_{(A)}^{\quad j}\) in 
complex co-ordinates by
\beer
\La_{(1)}^{\quad j} &\equiv& \la_{(1)}^{\quad j} \nn\\
\La_{(2)}^{\quad j} &\equiv& \dv{1}{\sqrt{2}}(\la_{(2)}^{\quad j} + 
  i\la_{(3)}^{\quad j}) \nn\\
\La_{(\tb)}^{\quad j} &\equiv& \dv{1}{\sqrt{2}}(\la_{(2)}^{\quad j}-
  i\la_{(3)}^{\quad j}), \nn
\eer
where complex frame indices take the values 1, 2, $\tb$.

We choose complex co-ordinates on the equipotential surfaces so that 
\[
\Lambda^j_{(2)}=\Sigma\delta^j_2 \qquad\qquad \Lambda^j_{(\tb)}=\overline{\Sigma}\delta^j_{\tb},
\]
for some comlex function $\Sigma$.

For a Weyl-type electric field the gravitational and electric potentials are related by
\[
e^{\om}  =  B + C\phi + \phi^2 \qquad B,C \;constants. 
\]
Assembling the field equations (\ref{feq1}) - (\ref{max}) (using (\ref{e:3driem}) and (\ref{aa1}) together with the commutation relations (\ref{a1})) gives
\beer
& &(\ln\si),_1  = -HU\label{b1}\\
& &UH,_1  + (HU)^2  + {\dv{1}{4}}\left(1 + \frac{e^{x^1}}{\lambda}\right)^{-1}  = 0\label{b2} \\
& &H,_2  = 0 \label{b3}\\
& &\Sigma\overline{\si},_2  + \overline{\Sigma}\si,_{\tb}  + 2|\si|^2  + H^2 -\dv{1}{4}U^{-2}
  \left(1 + \frac{e^{x^1}}{\lambda}\right)^{-1}  = 0 \label{b4}\\
& &(\ln U),_1 -2HU  + \frac{e^{x^1}}{2\lambda}\left(1 + \frac{e^{x^1}}{\lambda}\right)^{-1}  = 0\label{b5} \\
& &U,_2  = 0\label{b6} \\
& &(\ln\Sigma),_1  = -HU \label{b7}\\
& &\overline{\Sigma}\Sigma,_{\tb}  +  \overline{\si}\Sigma  = 0, \label{b8}
\eer
where
\[
\lambda \equiv {\frac{C^2}{4}} - B.
\]

By (\ref{b3}), (\ref{b6}) we have that $U = U(x^1$) and $H = H(x^1$), and so (\ref{b2}) and (\ref{b5}) are ordinary differential equations. In order to integrate this system 
of equations, we  must consider separately the cases $\lambda = 0$, $\lambda>0$ and 
$\lambda<0$. In the asymptotically flat case these correspond to $m = |e|$, $m>|e|$ and $m<|e|$, respectively.

\subsection*{IVa. $\lambda = 0$}

Here, (\ref{b2}), (\ref{b4}) and (\ref{b5}) reduce to
 \beer
& &UH,_1  + (HU)^2  = 0 \label{b9}\\
& &HU,_1 -2(HU)^2  + \hf HU  = 0\label{b10} \\
& &\overline{\Sigma}\si,_{\tb}  + \Sigma\overline{\si},_2  + 2|\si|^2  +  H^2  = 0. \label{b11}
\eer
 Adding the first two of these we get the Bernoulli equation
\[\frac{d(HU)}{dx^1}  + \hf HU  =  (HU)^2, \]
with solution
 \bee\label{b12}
HU  = \frac{e^{-\shf x^1}}{a + 2e^{-\shf x^1}},
\ee
where $a$ is an arbitrary constant of integration.

Now, (\ref{b12}) in (\ref{b10}) reads
\[(\ln U),_1 -\frac{2e^{-\shf x^1}}{a + 2e^{-\shf x^1}}  + \hf  = 0, \]
or, integrating this up 
 \bee\label{b13}
U = \frac{be^{-\shf x^1}}{(a + 2e^{-\shf x^1})^2}
\ee
with $b$ another arbitrary (non-zero) constant of integration. When this is put 
in (\ref{b12}) we find that
 \bee\label{b14}
H = \dv{1}{b}(a + 2e^{-\shf x^1})
\ee

Now, having determined the $\overline{g}_{11}$ component of the metric we return to 
equation (\ref{b7}), which, using (\ref{b12}), integrates up to 
\[\Sigma  = (a + 2e^{-\shf x^1})e^{S + is},\]
where $S$ and $s$ are otherwise arbitrary real-valued functions of ($z^2,z^{\tb}$).
 However, $s$ does not contribute to the metric form and so we 
can, without loss of generality, put $s = 0$. Thus
 \bee\label{b15}
\Sigma = (a + 2e^{-\shf x^1})e^S, 
\ee
and we need only determine $S$ to completely solve the problem.

We do this by first putting (\ref{b15}) in (\ref{b8}) yielding
 \bee\label{b16}
\si  = -(a + 2e^{-\shf x^1})S,_2e^S
\ee
and then equations (\ref{b14}) - (\ref{b16}) in (\ref{b11}) boils down to the simple equation 
 \bee\label{b17}
S,_{2\tb}  =  \frac{e^{-2S}}{2b^2}
\ee
In real co-ordinates this is the 2-dimensional Poisson-type equation
\[\frac{\partial^2S}{\partial x^2\:^2} + \frac{\partial^2S}{\partial x^3\:^2} 
   = \frac{e^{-2S}}{2b^2}. \]

The general solution of (\ref{b17}) is (see Nehari \cite{neh})
 \bee\label{b18}
e^S  =  \frac{1 + |\psi|^2}{\sqrt{2}|b||\psi,_2|},
\ee
where $\psi$ is an otherwise arbitrary function of $z^2$. Finally, (\ref{b1}) is 
identically satisfied by (\ref{b12}) and (\ref{b16}).

We can now assemble the metric with (\ref{b13}), (\ref{b15}) and (\ref{b18})
\[ds^2  =  b^2e^{-x^1}\left[\frac{e^{-x^1}}{(a + 2e^{-\shf x^1})^4}(dx^1)^2  + 
  \frac{4|\psi,_2|^2|dz^2|^2}{(a + 2e^{-\shf x^1})^2(1 + |\psi|^2)^2}\right] -
  e^{x^1}dt^2 ,\]
or making the co-ordinate transformation
 \bee\label{bb1}
x^1 \gt x'^1 \equiv x^1 \qquad z^2 \gt z'^2 \equiv 2\psi \qquad
  z^{\tb} \gt z'^{\tb} \equiv 2\overline{\psi},
\ee
and subsequently dropping the primes and returning to real co-ordinates
 \bee\label{b19}
ds^2  = b^2e^{-x^1}\left[\frac{e^{-x^1}}{(a + 2e^{-\shf x^1})^4}(dx^1)^2  + 
  \frac{1}{(a + 2e^{-\shf x^1})^2}\frac{(dx^2)^2 + (dx^3)^2}{(1 + \dv{1}{4}
  [(x^2)^2 + (x^3)^2])^2}\right] -e^{x^1}dt^2.
\ee
By writing this metric on a null tetrad and analyzing it with the REDUCE computer algebra system we find that this metric generates a 
Petrov type D field, as expected.  Furthermore, the surfaces $x^1 = constant$ in (\ref{b19}) are 2-spaces of constant positive curvature i.e. spheres and so we expect, by 
Birkhoff's Theorem, that (\ref{b19}) is just the extreme Reissner-Nordstr\"{o}m 
solution in an unusual co-ordinate system. To see that this is indeed the case we 
introduce new co-ordinates ($\theta,\varphi$) on these 2-spaces, defined by
 \bee\label{b20}
\theta \equiv tan^{-1}\left[\hf((x^2)^2 + (x^3)^2)^{\shf}\right] \qquad
 \varphi \equiv \tan^{-1}\left(\frac{x^3}{x^2}\right),
\ee
so that (\ref{b19}) reduces to
\[ds^2  = b^2e^{-x^1}\left[\frac{e^{-x^1}}{(a + 2e^{-\shf x^1})^4}(dx^1)^2  + 
  \frac{d\theta^2  + Sin^2\theta d\varphi^2}{(a + 2e^{-\shf x^1})^2}\right] - 
  e^{x^1}dt^2 .\]

Now, if we relabel the equipotential surfaces with $\varrho$ defined by 
\[ e^{-\shf x^1}  =  -\frac{a}{2}\left(1 + \frac{b}{2\varrho}\right), \]
and rescale the time co-ordinate by
\[t \gt t'\equiv \frac{2t}{a}, \]
we find, dropping the primes, that (\ref{b19}) becomes
\[ds^2  = \left(1 + \frac{b}{2\varrho}\right)^2\left[d\varrho^2  + 
  \varrho^2(d\theta^2  + Sin^2\theta d\varphi^2)\right]-
  \left(1 + \frac{b}{2\varrho}\right)^{-2}dt^2, \]
which is precisely the extreme Reissner-Nordstr\"{o}m solution with
\[b = 2m = 2|e|.\]

\subsection*{IVb. $\lambda> 0$}

In this case we must solve (\ref{b1})-(\ref{b8}). We still have, by (\ref{b3}) and (\ref{b6}) $U = U(x^1$) and $H = H(x^1$), but when we add (\ref{b2}) and (\ref{b5}) we get the Ricatti equation
 \bee\label{c1}
\frac{d(HU)}{dx^1} -(HU)^2  + \hf HU\dv{e^{x^1}}{\lambda} \left(1 + \frac{e^{x^1}}{\lambda}\right)  ^{-1}  + 
  \dv{1}{4}\left(1 + \frac{e^{x^1}}{\lambda}\right) ^{-1}  = 0,
\ee
with first integral
 \[
HU = \hf\left(1 + \frac{e^{x^1}}{\lambda}\right) ^{-\shf}.
\]

So returning to (\ref{c1}) with a substitution of the form
 \bee\label{c2}
HU  = \hf\left(1 + \frac{e^{x^1}}{\lambda}\right) ^{-\shf}  +  F(x^1),
\ee
we get a Bernoulli equation for $F$ with solution
 \bee\label{c3}
F  = \frac{ 1}{ae^{-I}-1}\left(1 + \frac{e^{x^1}}{\lambda}\right) ^{-\shf},
\ee
where $a$ is an arbitrary constant of integration and $I$ is defined
 \[
I \equiv \int\left(1 + \frac{e^{x^1}}{\lambda}\right) ^{-\shf}dx^1.
\]

Plugging (\ref{c3}) in (\ref{c2}) gives the general solution to (\ref{c1}) as
 \beer
HU& = &\hf\left(1 + \frac{e^{x^1}}{\lambda}\right) ^{-\hf}  + \frac{e^I }{a-e^I}\left(1 + \frac{e^{x^1}}{\lambda}\right) ^{-\shf}  \nn\\
& = &{\hf}\left(\frac{ae^{-I} + 1}{ae^{-I}-1}\right)\left(1 + \frac{e^{x^1}}{\lambda}\right) ^{-\shf}.\label{c4}
\eer

 Putting this in (\ref{b5}) and integrating
 \bee\label{c5}
U = \frac{be^I}{\left(1 + \frac{e^{x^1}}{\lambda}\right) ^{\shf}(a-e^I)^2},
\ee
where $b$ is a non-zero constant of integration. Then (\ref{c5}) in (\ref{c4}) gives
 \bee\label{c6}
H = \dv{1}{2b}e^{-I}(a^2-e^{2I}).
\ee

Again, having determined \(\overline{g}_{11}\) we return to (\ref{b7}) which, with 
the help of (\ref{c5}) and (\ref{c6}) integrates to
 \bee\label{c7}
\Sigma  =  (a-e^I)e^{-\shf I}e^{S + is},
\ee
where $S$ and $s$ are arbitrary functions of ($z^2,z^{\tb}$). Without loss of 
generality, we can again put $s = 0$.

Then, (\ref{c7}) in (\ref{b8}) gives
 \bee\label{c8}
\si  = -(a-e^I)e^{-\shf I}S,_2e^S,
\ee
and (\ref{b1}) is identically satisfied by (\ref{c7}) and (\ref{c8}).

Our remaining equation, (\ref{b4}) boils down, from (\ref{c5}) - (\ref{c8}), to
 \bee\label{c9}
S,_{2\tb}  =  \frac{a}{2b^2}e^{-2S}.
\ee

Here we must consider the solutions to (\ref{c9}) for $a>0$, $a = 0$ and $a<0$ 
separately.  These correspond to the cases where the equipotential surfaces are of constant positive, zero and negative curvature, respectively.

 \begin{description}
\item{$\ul{a>0}$:}

In this case the general solution to (\ref{c9}) is
\[e^S  = \sqrt{\dv{a}{2b^2}}\left[\frac{1 + |\psi|^2}{|\psi,_2|}\right], \]
where $\psi$ is an otherwise arbitrary function of $z^2$.

Assembling our metric we find that
\[ds^2 = b^2e^{-x^1}\left[\frac{e^{2I}(dx^1)^2}{\left(1 + \frac{e^{x^1}}{\lambda}\right) (a-e^I)^4}  + 
  \frac{4e^I|\psi,_2|^2|dz^2|^2}{a(a-e^I)^2(1 + |\psi|^2)^2}\right] -
  e^{x^1}dt^2 ,\]
and then transforming co-ordinates by (\ref{bb1}) and dropping the primes we find
\[ds^2  = b^2e^{-x^1}\left[\frac{e^{2I}(dx^1)^2}{\left(1 + \frac{e^{x^1}}{\lambda}\right) (a-e^I)^4}  + 
 \frac{e^I((dx^2)^2 + (dx^3)^2)}{a(a-e^I)^2[1 + \sdv{1}{4}((x^2)^2 + (x^3)^2)]^2}
 \right] -e^{x^1}dt^2. \]
Again, the field is Petrov type D, the equipotential surfaces are 2-spaces of 
constant positive curvature i.e. spheres, and we thus expect it to represent the 
under-charged ($m>|e|$) or over-charged ($m<|e|$) Reissner-Nordstr\"{o}m solution in an unfamiliar 
co-ordinate system. This is indeed the case, as can be seen by utilizing the 
co-ordinate transformation (\ref{b20}) and relabeling the equipotential surfaces with 
$\varrho$ defined by
\[ e^I \equiv a\left(\frac{b-\varrho}{b + \varrho}\right),\]
or, equivalently
\[e^{x^1} \equiv \frac{4\lambda a}{(1-a)^2}\frac{(1-\frac{b^2}{\varrho^2})^2}
  {\left[1 + 2\left(\frac{1 + a}{1-a}\right)\frac{b}{\varrho}  + \frac{b^2}{\varrho^2}
  \right]^2}. \]
The metric then becomes
\[ds^2 = \frac{(1-a)^2}{64a^3}\left[1 + 2\left(\frac{1 + a}{1-a}\right)\frac{b}
{\varrho} + \frac{b^2} {\varrho^2}\right]^2\left[d\varrho^2  + \varrho^2(d\theta^2
  + Sin^2\theta d\varphi^2)\right] \]
\[-\frac{4\lambda a(1-\frac{b^2}{\varrho^2})^2 }{(1-a)^2}\left[1 + 2
  \left(\frac{1 + a}{1-a}\right)\frac{b}{\varrho}  + \frac{b^2}{\varrho^2}\right]
^{-2} dt^2, \]
which, with a final rescaling of
\[t \gt \left(\frac{1-a}{2\sqrt{\lambda a}}\right)t \qquad \varrho \gt 
  \left(\frac{(4a)^{\sdv{3}{2}}}{1-a}\right)\varrho, \]
takes the standard isotropic form
\[ds^2 = \left[1 + 2\left(\frac{1 + a}{1-a}\right)\frac{b}{\varrho}  + \frac{b^2}
{\varrho^2} \right]^2\left[d\varrho^2  + \varrho^2(d\theta^2  + Sin^2\theta
 d\varphi^2) \right]\]
\[-\left(1-\frac{b^2}{\varrho^2}\right)^2\left[1 + 2\left(\frac{1 + a}{1-a}
  \right)\frac{b}{\varrho}  + \frac{b^2}{\varrho^2}\right]^{-2}dt^2 ,\]
with mass and charge parameters given by
\[m = \frac{2(1 + a)b}{1-a} \qquad e^2  = \frac{16ab^2}{(1-a)^2}. \]

\item{$\ul{a = 0}:$}

In this case the general solution to (\ref{c9}) is
\[ e^{-S}  = b|\psi,_2| \qquad \psi  = \psi(z^2), \]
and using (\ref{bb1}) our metric takes the simple form
\[ds^2 = b^2e^{-x^1}\left[\frac{e^{-2I}}{\left(1 + \frac{e^{x^1}}{\lambda}\right) }(dx^1)^2  + 
  e^{-I}\left[(dx^2)^2 + (dx^3)^2\right]\right] -e^{x^1}dt^2. \]

From this we can see that the equipotential surfaces are flat and so our 
metric represents the field of a charged infinite plate which also turns out 
to be of Petrov type D. The fact that it is not an asymptotically flat field 
can be most easily seen by transforming into isotropic co-ordinates. However, 
rather than do this here, we shall change to the canonical form of the plane 
symmetric static Einstein-Maxwell field, as given by McVitie \cite{mcv}. This is 
achieved by relabelling the equipotential surfaces 
by $z$ defined by
\[ e^{x^1} \equiv \frac{4(\lambda^2 +\lambda)^{\shf}}{bz}  + \frac{4(1 + \lambda)}{b^2z^2}, \]
and rescaling
\[t \gt t'\equiv \left(\frac{\lambda^{\shf}b^2}{(\lambda + 1)^{\shf}}\right)t  \]
\[x^2 \gt x'^2\equiv \left(\frac{b^2}{(\lambda + 1)^{\shf}}\right)x^2 \qquad
x^3 \gt x'^3\equiv \left(\frac{b^2}{(\lambda + 1)^{\shf}}\right)x^3, \]
we get (dropping the primes) 
\[ds^2  = \left(\frac{m}{z}  + \frac{e^2}{z^2}\right)^{-1}dz^2  + z^2\left[(dx^2)^2  + 
  (dx^3)^2\right] - \left(\frac{m}{z}  + \frac{e^2}{z^2}\right)dt^2, \]
where
\[ m \equiv \frac{4(\lambda + 1)^{\dv{3}{2}}}{\lambda^{\shf}b^5} \qquad e^2 \equiv \frac
  {4(\lambda + 1)^2}{\lambda b^6}. \]

\item{$\ul{a<0}:$}

In this case (\ref{c9}) has general solution
\[ e^S = \sqrt{\frac{-a}{2b^2}}\left[\frac{1-|\psi|^2}{|\psi,_2|}\right], \]
and so, assembling our metric we find that, after transforming by (\ref{bb1}) that
\[ d\overline{s}^2 = d\left(\frac{1}{e^I-a}\right)^2-\frac{e^I}{a(e^I-a)^2}\frac{
  (dx^2)^2 + (dx^3)^2}{\left[1-\dv{1}{4}[(x^2)^2 + (x^3)^2]\right]^2}. \] 

If we write the metric for the equipotential 2-surfaces of constant negative 
curvature in the form Stephani \cite{steph}
\[ d\overline{\si}^2 = \frac{du^2}{1 + u^2} + u^2d\varphi^2, \]
and make the co-ordinate transformations
 \beer
x^1&\gt&\varrho \equiv \left[\frac{-e^I}{a(e^I-a)^2}\right]^{\shf}u \nn\\
   \noalign{\vskip 6pt}
u&\gt&h \equiv (1 + u^2)^{\shf}\left[\frac{1}{4a^2}-\frac{\varrho^2}{u^2}
   \right]^{\shf} ,\nn
\eer
we find that the metric is in the Weyl canonical form for a static 
axially symmetric field. We furthermore find that the field is singular 
along the axis of symmetry and has only two principal null directions 
associated with the Weyl curvature tensor (i.e. it is Petrov type D).
We therefore interpret the field as being generated by an infinite rod 
situated along the axis of symmetry $\varrho = 0$.

\end{description}

\subsection*{IVc. $\lambda< 0$}
By a series of calculations similar to the $\lambda>0$ case, one obtains the metric generated by an overcharged ($m<|e|$) sphere, infinite rod and infinite plate.

\section*{V Shearing Fields }

By a procedure similar to that of the last section all Weyl-type fields with shearing geodesic lines of force can be found. The metric $\overline{g}$ turns out to be (for details, see Guilfoyle \cite{guil})
\beer
\overline{g}_{11}& = &\frac{F^{-2}}{\la^2(\la + e^{x^1})}\left[1 + 2k\la F[\dv{1}{4}((x^2)^2 + 
  (x^3)^2)^2 + [(x^2)^2-(x^3)^2]\Omega  + \Omega^2]\right] \nn\\
\overline{g}_{12}& = &-\frac{F^{-1}x^2}{\la(\la + e^{x^1})^{\shf}}[\hf((x^2)^2-3(x^3)^2) + 
  \Omega] \nn\\
\overline{g}_{13}& = &-\frac{F^{-1}x^3}{\la(\la + e^{x^1})^{\shf}}[\hf((x^3)^2-3(x^2)^2)- 
  \Omega] \nn\\
\overline{g}_{22}& = &\overline{g}_{33} = \frac{1}{k\la}\left((x^2)^2 + (x^3)^2\right)F^{-1}, \nn
\eer
where $k$ is a constant, $\Omega$ is an arbitrary function of $x^1$, and
\beer
F(x^1) & = & exp\left[ (\la  + 4k^2)^{\shf} \int (\la  + e^{x^1})^{-\shf}dx^1\right] \nn\\
                \noalign{\vskip 6pt}
    & = & \left\{ \begin{array}{ll}
            \left[\frac{\left(\la + e^{x^1}\right)^{\shf}- \la^{\shf}}{\left(\la + e^{x^1}\right)^{\shf} + \la^{\shf}}\right]^n  &  for \;\la>0  \nn\\
\noalign{\vskip 6pt}
                \exp\left[-2n\tan^{-1}\left[-\left(1 + \dv{e^{x^1}}{\la}\right)\right]^{\shf}\right]  &  for \;\la<0,
                \end{array} \right. 
\eer
for \(n \equiv [1 + \dv{(2k)^2}{\la}]^{\shf} \).  As expected, this metric is algebraically general, with shear given by
\[|\beta|^2 = k^2\la^2F^2\not=0 .\]
Therefore this cannot be of Majumdar-Papapetrou type. That is $\lambda>0$ or $\lambda<0$, but $\lambda = 0$ is not possible.

The equipotential surfaces $x^1=constant$ are easily seen to be flat.

\section*{VI. Conclusion}

We have established that all algebraically special Weyl-type electric fields have shearfree geodesic lines of force.  In the over-charged ($m<|e|$) and under-charged ($m>|e|$) cases these are the fields generated by a charged sphere, infinite charged rod and infinite charged plate. These correspond exactly to the fields in Newtonian theory that have straight lines of force. Of these, only the sphere is asymptotically flat.  In the Majumdar-Papapetrou case ($m=|e|$) the only solution is the extreme Reissner-Nordstr\"{o}m solution.  All metrics with shearing geodesic lines of force giving rise to a Weyl-type electric field were also found.  This class of solutions depends on an arbitrary function of the gravitational potential and the equipotential surfaces are flat.  It is algebraically general and cannot be in the Majumdar-Papapetrou class.

\section*{VII. Acknowledgment}

The author would like to thank Petros Florides, under whose supervision most of the above work was carried out.

\end{document}